\documentclass{kluwer}
\begin{document}
\begin{article}
\begin{opening}
\title{Cosmology with Supernovae\texttt{} \thanks{Invited Review
given at JENAM 2002, Porto 4-6 September}}            
\subtitle{The Road from Galactic Supernovae to the edge of the Universe}

\author{P. \surname{Ruiz--Lapuente}} 
\institute{Department of Astronomy, U. Barcelona\\
 CER for Astrophysics, Particle Physics and Cosmology, U. Barcelona\\ 
Max--Planck
Institut f\"ur Astrophysik, Garching bei M\"unchen}                               




\runningtitle{Cosmology with Supernovae}
\runningauthor{Ruiz--Lapuente}



\begin{abstract} 
This review gives an update of the cosmological use of 
SNe Ia and the progress made in 
testing their properties from the local universe to high--z.
The cosmological road from high--z supernovae 
down to  Galactic SNe Ia is followed in search of the answer to standing
questions on their nature and their validity as cosmological 
indicators. 

\end{abstract}




\end{opening}

\section{Introduction}
  
The use of SNe Ia as {\it calibrated candles} has recently led to a fundamental
discovery: that the rate of the cosmic expansion of the Universe is 
accelerating (Perlmutter et al. 1999; Riess et al. 1998).
The presence of a energy component with negative pressure 
(still undistinguishable from the cosmological constant $\Lambda$) and 
 the nature of this new 
component is a major challenge in Cosmology and in Fundamental Physics. 
Precise enough observations of SNe Ia extending to redshifts $z\sim 1.5-2.0$ 
should yield the {\it equation of state} of the new component (usually termed 
{\it dark energy}), $p_{X} = w_{X} \rho_{X}$, that is the relationship 
between pressure and energy density.

 Both the new picture of the Universe that is
now emerging and the next step in its investigation, that of determining 
the nature of {\it dark energy},  critically depend on
the reliability of the calibration of SNe Ia luminosities, which up to now is 
purely empirical. While various efforts to build up 
large consistent samples of SNe Ia covering the z range needed
for cosmology are on their way, detailed analyses of the current
high--z sample have shed new light into old questions.

\section{The Use of Type Ia supernovae for cosmology}

\subsection{A brief historical account}

Type Ia supernovae (SNe Ia) are explosions of carbon--oxygen white dwarfs 
(C+O WDs), very uniform in their physical 
properties. Their brightness and  the possibility 
of calibrating their luminosity makes them very suitable distance indicators 
in cosmology. 

In the first uses of SNe Ia to determine the cosmological parameters it was 
assumed that those objects were standard candles, implying that they all had 
similar luminosities. In 1977, Pskovskii realized that there
was a correlation between the brightness at maximum and the rate of decline 
of the light curve: the brighter SNe Ia have a slower decline of their light 
curves whereas fainter ones would be faster decliners.
A systematic follow--up of 
SNe Ia (Maza et al. 1994; Filippenko et al 1992a,b;
 Leibundgut 1993; Phillips 1993; Hamuy et al 1996a,b; Barbon et al. 1999) 
confirmed the brightness--decline relation. 
Phillips (1993), and Hamuy et al. (1996a,b)                                    
quantified it by studying supernovae at $z\sim 0.1$, a z large enough that 
peculiar motions do not introduce dispersion in the magnitude--redshift 
diagram. The intrinsic variation of SNe Ia is written as a linear relationship 
of the sort:   

\begin{equation}
M_{B} = -M_{Bt} + \alpha  \ [\Delta m_{15}(B) - \beta] + 5 \ log (H_{0} / 65) 
\end{equation}  

\noindent
$\Delta m_{15}$, the parameter of the SNe Ia light curve family, is the 
number of magnitudes of decline in 15 days after maximum. The value of 
$\alpha$ as well as the dispersion of that relation have been evaluated in 
samples obtained from 1993 to the present (see Phillips et al. 
1999). $M_{Bt}$ is the absolute magnitude in $\rm{B}$ for a 
template SNIa of $\beta$ rate of decline.

\begin{figure}
\input epsf2
\bigskip
\bigskip
\bigskip
\epsfxsize=210pt
\epsfbox{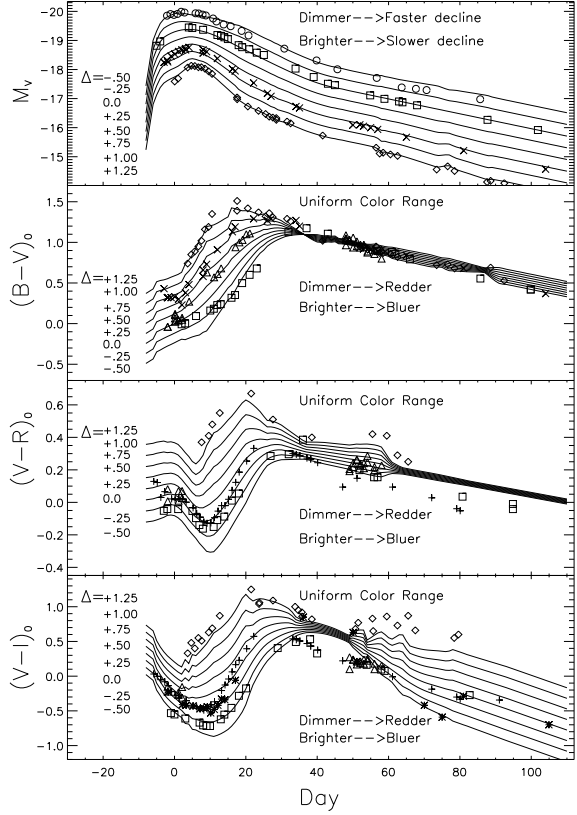}
\vspace{6mm}
\caption{
 The correlation maximum brightness--rate of decline of SNe Ia as 
described throught the SNe Ia light curve shapes by Riess, Press \& 
Kirshner (1995a).                
}
\end{figure}

\begin{figure}
\input epsf2
\bigskip
\bigskip
\bigskip
\epsfxsize=210pt
\epsfbox{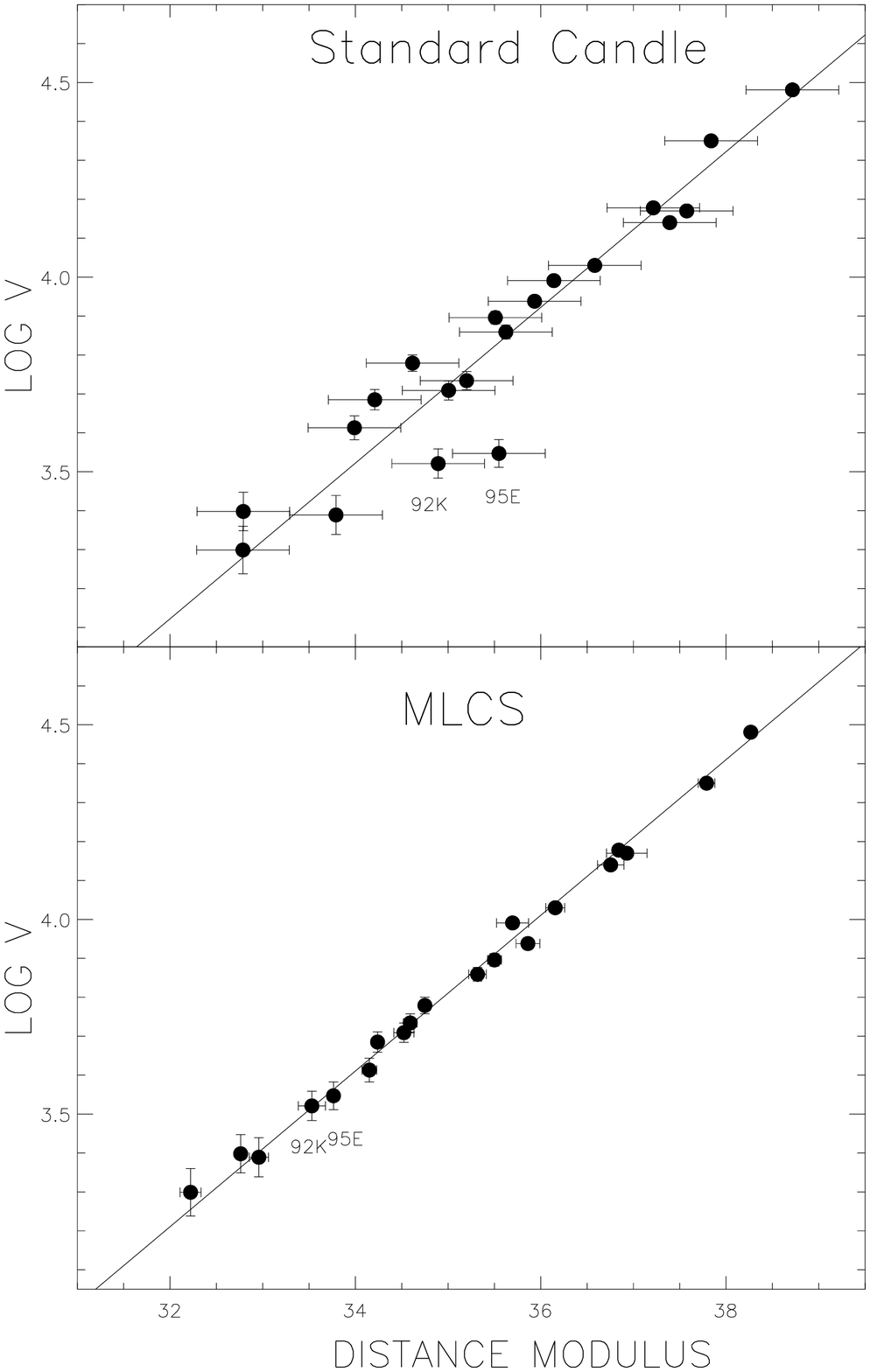}
\vspace{6mm}
\caption{
  The Hubble diagram for SNe Ia before (top panel) and after (bottom 
panel) correction of the width--absolute brightness relationship. The figure 
is from Riess, Press \& Kirshner (1995a).              
}
\end{figure}

Other groups have formulated the brightness--decline correlation in a 
different way. Riess, Press \& Kirshner (1995a,b)
use the full shape of the light curve with respect to a template. This is 
also the formulation used by the High--Z SN Team in their high--redshift 
supernova studies. The Supernova Cosmology Project collaboration, on the 
other hand, has introduced the {\it stretch--factor, s} as a parameter to 
account for the brightness--decline 
relationship (Perlmutter et al. 1999, Goldhaber et al. 2001).

All different ways to parameterize the effect lead to equivalent 
results for the present value of the rate of expansion of the Universe 
H$_{0}$ and derive similar  matter density $\Omega_{M}$ and cosmological 
constant density $\Omega_{\Lambda}$. Leibundgut (see this conference)
shows that the corrections provided by different methods to the same 
supernovae are however different. By looking at the conversions 
between the MLCS, $\Delta m_{15}$ and {\it stretch factor}
 one sees that there are no  
linear relationships shifting one into the other and the approximate 
established conversions have some dispersion. While the overall results
on the cosmological implications from supernovae
are in agreement, it is important for the
 various methods  to describe in a complete 
reproducible way the procedure in which 
the brightness--rate of decline relationship
is measured.

The degree to which the brightness--rate of decline
 correlation reduces the scatter in the Hubble 
diagram is shown in Figure 2. On the other hand, the family of SNe Ia form a 
sequence of highly resembling spectra with subtle changes in some spectral 
features correlated with the light curves shapes (Nugent et al. 1998).

\begin{figure}
\input epsf2
\bigskip
\bigskip
\bigskip
\epsfxsize=210pt
\epsfbox{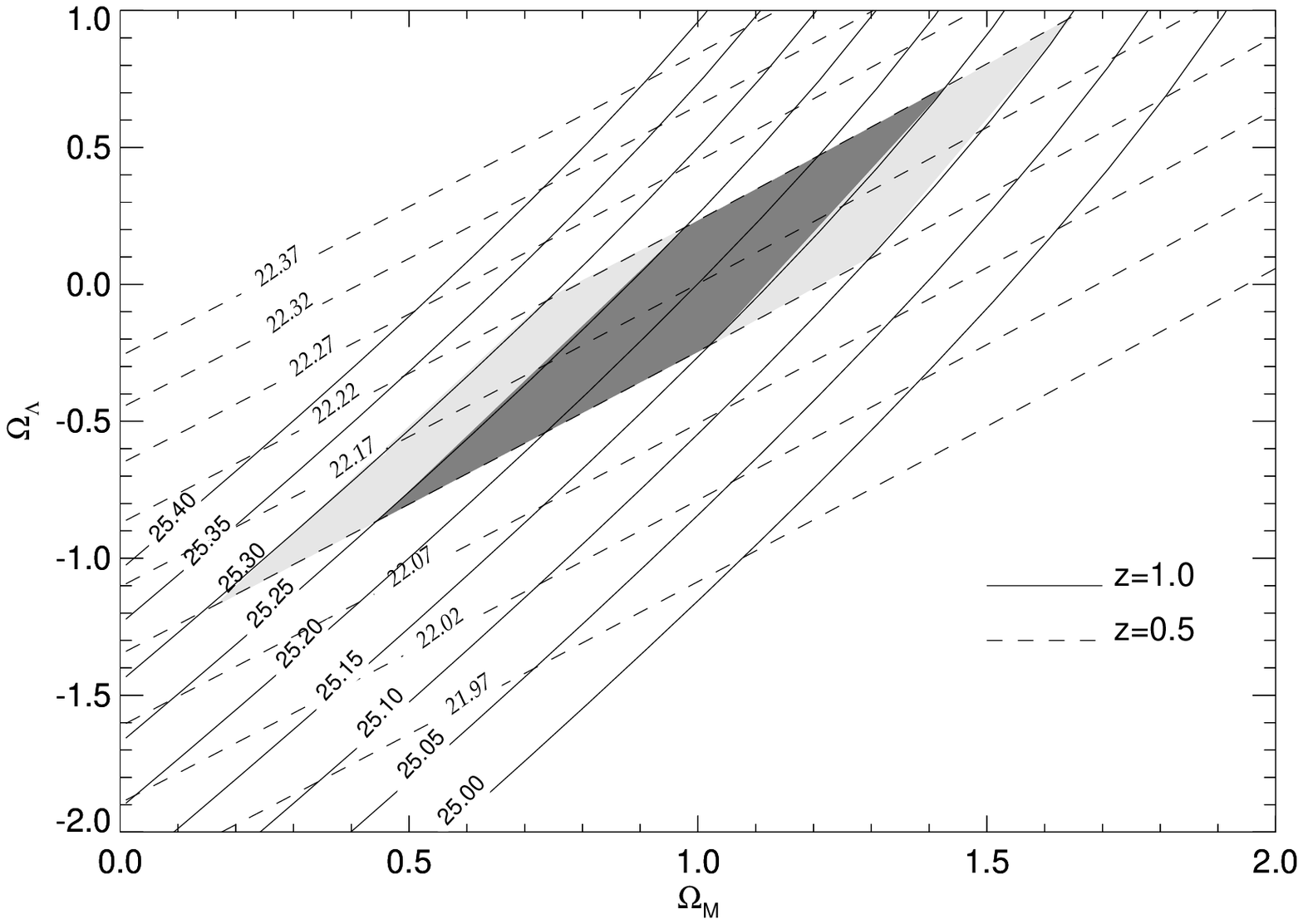}
\vspace{6mm}
\caption{
 Constraints in the $\Omega_{M}$--$\Omega_{\Lambda}$ plane from two 
SNe Ia (one at z =1.0 and another at  z = 0.5 according to Goobar 
\& Perlmutter (1995). SNe Ia at different z allow to 
discriminate between possible Universe models.
}
\end{figure}

By using the magnitude--redshift relation m(z) as a function of 
$\Omega_{M}$ and  $\Omega_{\Lambda}$ with a sample of high--z SNe Ia of 
different z, Goobar \& Perlmutter (1995) showed that it is 
possible to constrain, from observations, the region of allowed values in the 
$\Omega_{M}$--$\Omega_{\Lambda}$ plane. The relation goes as:
 
\begin{equation}
m(z) = M + 5\ log\ d_{L} (z, \Omega_{M}, \Omega_{\Lambda}) - 5\ log\ H_{0} +
K_{c} + 25
\end{equation}

\noindent
where M is the absolute magnitude of the supernova, $d_{L}$ the luminosity 
distance, and $K_{c}$ is the K--correction. The method is 
illustrated in Figure 3.

By 1998, such use gave important results in the two collaborations. 
Those results were presented in the $\Omega_{M}$--$\Omega_{\Lambda}$ plane 
and implied $\Omega_{\Lambda} > 0$  at a 3$\sigma$ confidence level. For a 
flat universe ($\Omega_{Tot} = 1$), the results from the Supernova Cosmology 
Project implied $\Omega_{M}$=$0.28^{+0.09}_{-0.08}$(stat)$^{+0.05}_{-0.04}$ 
(syst), and the High--Z Supernova Team obtained for a flat universe 
$\Omega_{M}$=0.24$\pm$0.1. The outcoming picture of our universe is that 
about 20--30$\%$ of its density content is in matter and 70--80\% in 
cosmological constant. According to the allowed  $\Omega_{M}$ and 
$\Omega_{\Lambda}$ values, the Universe will expand forever, accelerating its 
rate of expansion.

\subsection{Questions answered in the last years}

Two important questions were raised soon after the results from 1998 were 
known. The method  finds Type Ia supernovae at high $z$ 
being dimmer than what they should be for an Einstein--de Sitter universe or 
an open $\Lambda = 0$ Universe. Could just dust be responsible for that 
effect? (Aguirre 1999; Rowan--Robinson 2002).

A second important question relates to the universality of SNe Ia light 
curves. Do high-z and low--z SNe Ia  show the same brightness--rate of 
decline relationship? A difference in the rise time to maximum between SNe Ia 
at low and high $z$ was suggested in an early analysis by the High--Z SN 
Team (Riess et al. 2000). In general, one could  ask whether evolution and 
enviromental effects influence the observed SNe Ia light curves.

The most recent work has tried to clarify the above topics. 
Dust can be discriminated against a cosmological effect by going to a 
redshift beyond $z\sim 1$, where the predicted behavior of the curves differs 
significantly in each case, or/and by performing multicolor light curve 
analysis of the individual SNe Ia (Riess et al. 2000). 
SN 1997ff was a supernova in an epoch where the Universe was 
still decelerating and showed in the Hubble diagram that the effect
measured by the two collaborations was not linked to an overall dust
extinction that would cumulate with z (Riess et al. 2001).

In relation to the second question, that of the rise time to maximum of the 
supernova light curves, the study of a significant sample of SNe Ia at 
various $z$ by the Supernova Cosmology Project reveals that the rise times to 
maximum are similar in low--$z$ and high--$z$ 
SNe Ia (Aldering et al. 2000; Goldhaber et al. 2001).
 Statistical evaluation is so far consistent 
with no difference among the low--$z$ and high--$z$ samples. 

Moreover, the possibility of existence of an extra parameter in the maximum 
brightness--rate of decline relation has been carefully examined. Among 
possible influences, metallicity was one of the obvious ones to consider. We 
know that it plays an important role in the Cepheid period 
luminosity--relationship which is widely used in distance determinations at 
more modest redshifts.

\begin{figure}
\input epsf2
\bigskip
\bigskip
\bigskip
\epsfxsize=210pt
\epsfbox{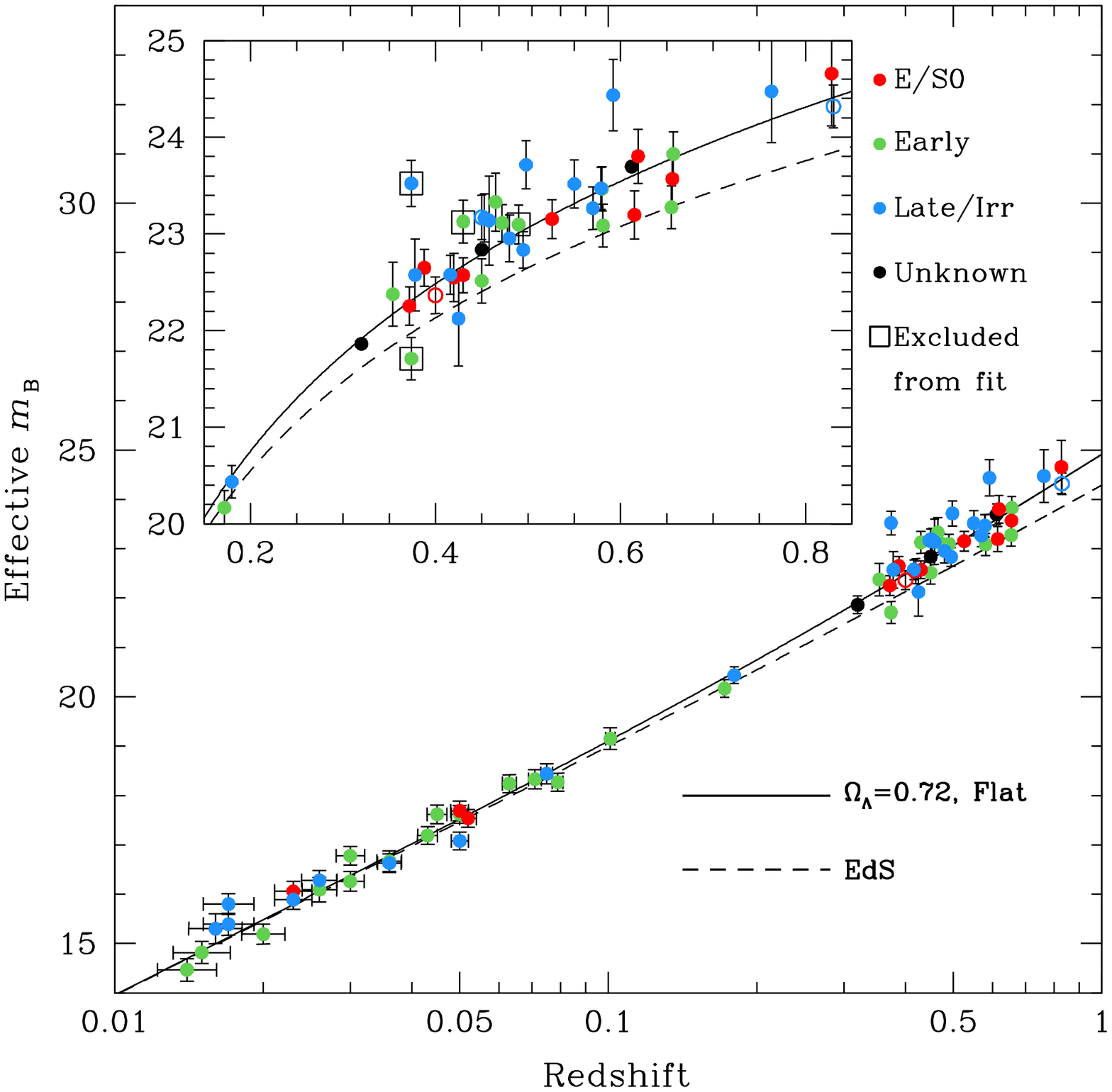}
\epsfxsize=210pt
\epsfbox{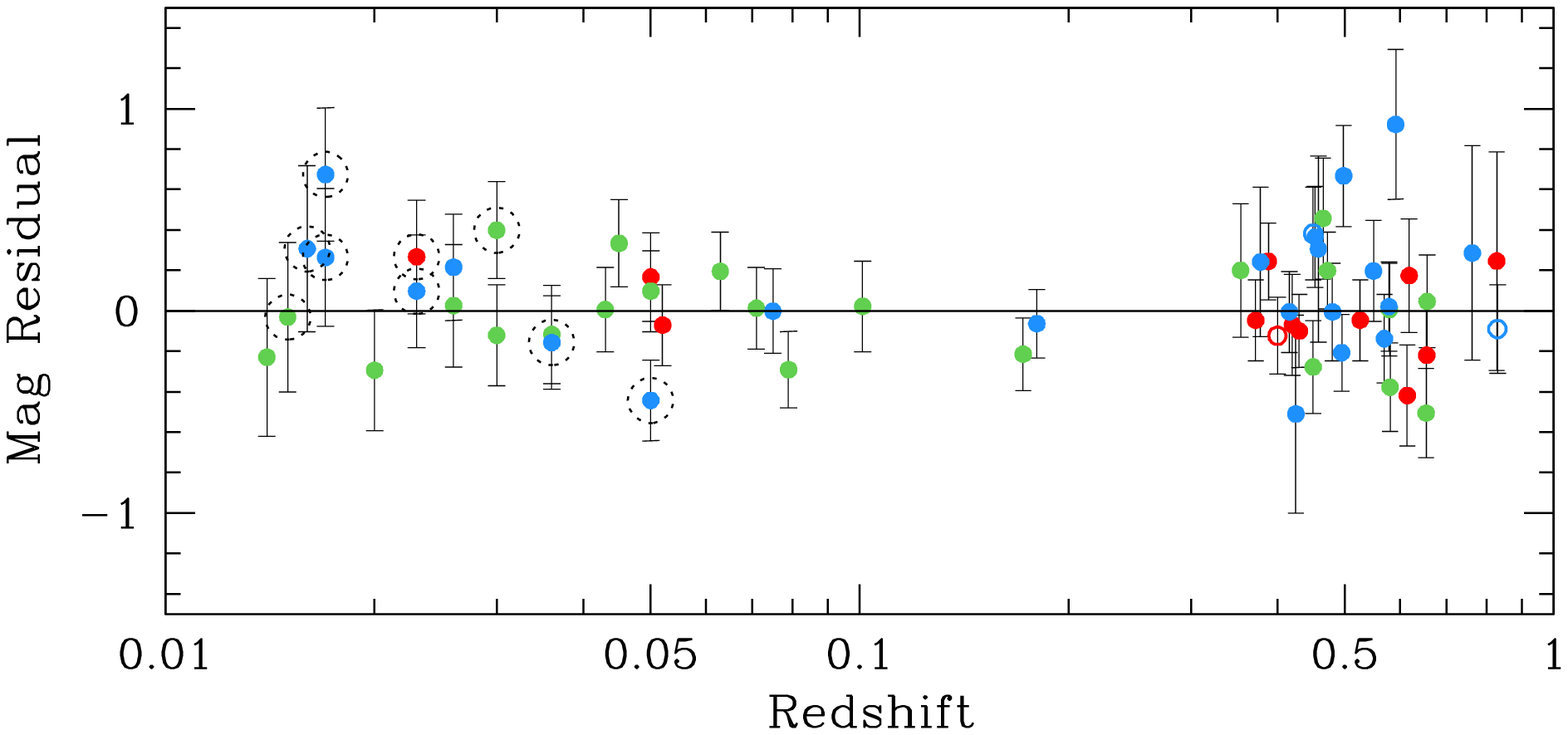}
\vspace{6mm}
\caption{Figure from Sullivan et al. (2002). 
Upper panel: The stretch--corrected SNe Ia Hubble diagram plotted 
according to the type of the host galaxy. Lower panel: Residuals 
from the adopted cosmology (`fit--C' of Perlmutter et al. 1999 
$\Omega_{M}=0.28$ and $\Omega_{\Lambda}=0.72$) for both
high and low redshift SNe. The residuals of SNe Ia in E/SO galaxies are
small.  
}
\end{figure}

Work by Ivanov, Hamuy \& Pinto (2000) has shed 
light on the issue. After examining supernovae in galaxies with a gradient of 
metallicity, they conclude that there is no evidence for metallicity 
dependence as an extra parameter in the light curve correlation of SNe Ia. 
Their result comes from an analysis of the SNe Ia light curves of
 a sample of 62 supernovae in the local Universe. The SNe Ia
 belong to different  
 populations along a metallicity gradient. 
This check has been done as well at high--z by Quimby et al. (2002) using
74 SNe Ia (0.17 $<$ z $<$ 0.86) from the SCP sample. No significant 
correlation between peak SNIa luminosity and metallicity is found.
The tendency to have dimmer 
(and faster declining) SNe Ia in elliptical galaxies than in 
spirals, encompassing a narrower distribution in the 
brightness--rate of decline relationship, seems to be a population age effect. 
In former discussions of the evolutionary path that leads to SNe Ia, 
Ruiz--Lapuente, Canal \& Burkert (1995, 
1997) argued for a 
population--age effect in the change in the dispersion of sample properties 
between ellipticals and spirals: less massive and colder WD SN progenitors 
would be selected in older populations. Spirals would contain WDs with a 
wider spread in masses and degrees of cooling than ellipticals.

\begin{figure}
\input epsf2
\bigskip
\bigskip
\bigskip
\epsfxsize=210pt
\epsfbox{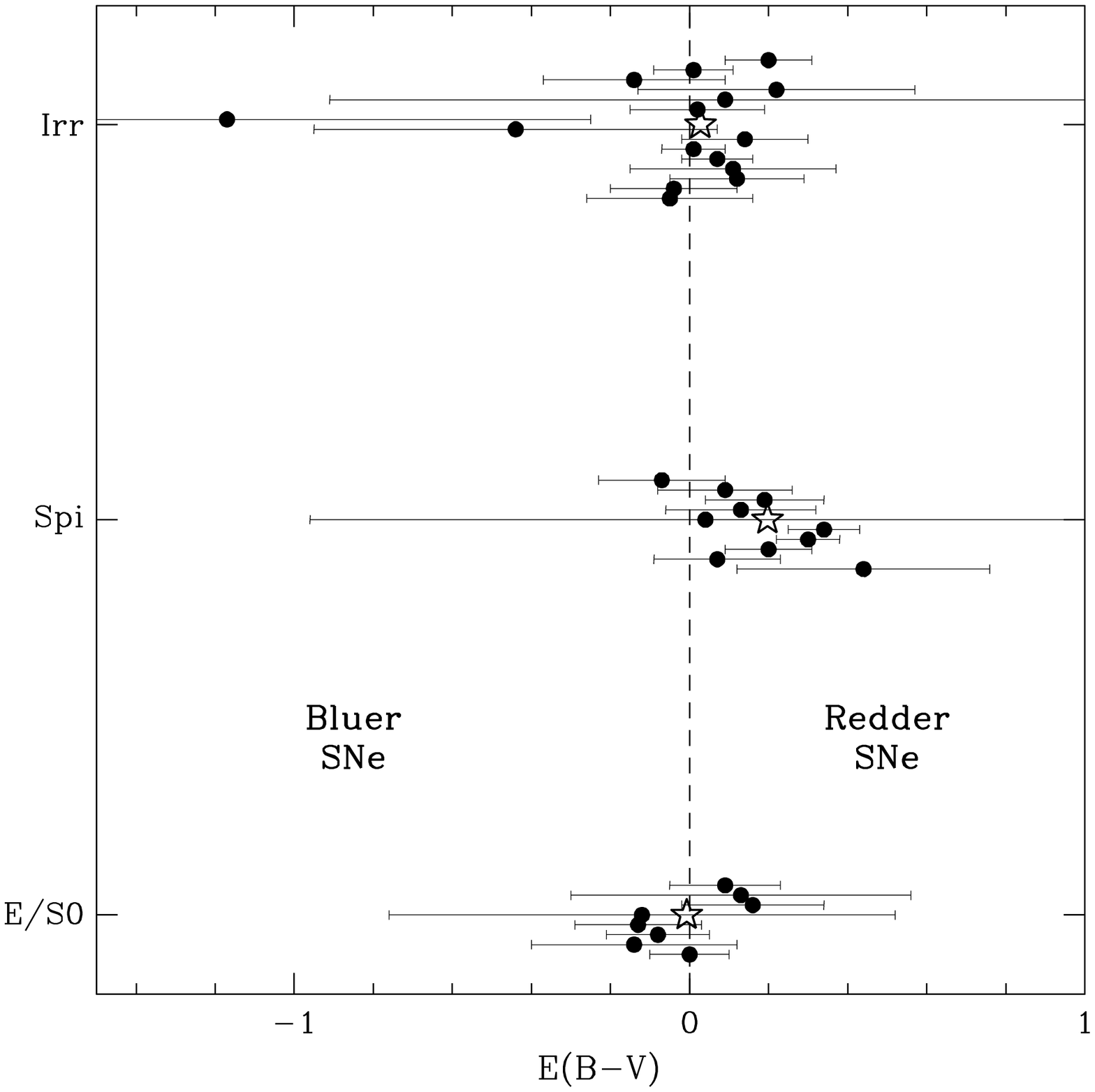}
\vspace{6mm}
\caption{The SN rest--frame ``colour 
excess'' E(B-V) plotted as a function of host galaxy type. Error bars are 
are derived from R and I photometry for each SN. Figure from
 Sullivan et al. (2002). 
}
\end{figure}

On the other hand, a study with the HST of
 the host galaxies of the SNe Ia sample of 
the Supernova Cosmology Project (Ellis \& Sullivan 2000; Sullivan et al. 2002)
suggests that, at high $z$, early--type galaxies do also show a narrower 
dispersion in SNe Ia properties than late--type galaxies, as they do at low 
$z$ (Branch, Baron \& Romanishin 1996).  
This study deserves special attention as it clarifies different points on
the link of SNe Ia properties with host galaxy type, the role of dust
and the shift in properties in the family of SNe Ia. Sullivan et al (2002)
examine the SCP (Perlmutter et al. 1999) sample according to its host galaxy
type. The SNe Ia in spiral galaxies appear fainter and redder and 
show a larger scatter around the best--fitting cosmological model than
the SNe Ia in E/S0 galaxies. The SNe Ia in dust--free galaxies like
 ellipticals show
a very reduced scatter (see Fig 4).  
The obvious explanation coming out of this work is that SNe Ia in late
galaxy types suffer from increased extinction in the galaxy. The 
SNe magnitudes are corrected regularly from Galactic extinction, but
the correction by dust residing in the host galaxy or along the line of sight
is not included in the standard fit, as it would require to have extensive 
color information that might not be available for high--z SNe Ia.
 An interesting discussion on the limits of dust extinction
as probed by this sample can be found in that paper, and it helps to
quantify the role of dust in making supernovae fainter.

Moreover, Sullivan et al. (2002) and Farrah et al. (2002) tested as well 
the sample of high-z SNe Ia along projected radius in their host
galaxies. Farrah et al. (2002) found no evidence that the 
SNe Ia at z=0.6 are preferentially found in outer regions ($>$ 10 kpc)
of host galaxies where extinction would be low. This suggests that the
 range of host galaxy extinctions of SNe Ia at z $\sim$ 0.6 is comparable
to that of local SNe Ia. No significant trends were found in stretch 
and other properties along projected radius between the low z and 
high z samples as well (Sullivan et al. 2002).

Another second interesting aspect examined in this work is the possibility
of a shift in the population distribution between SNe found at low and
 high redshift: at high redshift we see in the Permutter et al.(1999)
sample no trend in stretch with galaxy type, whereas at low redshift 
SNe with larger stretches (slow decliners) 
are found in later--type galaxies and are missing
from  {\it E/S0} galaxies (though this is a tentative result given the
size of the high--z  sample).

The reader will note that here we are talking about the spread of the SNe Ia 
samples in galaxies of different types: on how many fast SNe Ia versus 
intermediate or slow decliners are found in the various morphological types. 
Spirals at low or moderate redshifts should encompass all ranges of variation 
in the SNe Ia properties since they contain populations with a wide spread in 
age. The cosmological collaborations keep finding that in those samples the 
one--parameter correlation does give a good description of the variation of 
the SNe Ia light curves (Riess et al. 2000; Goldhaber et al. 2001).
   
The fact that in old populations SNe Ia are systematically dimmer than in 
young populations could be interpreted in terms of what affects the 
progenitors of the SNeIa: the WDs. A population age effect that we 
favor (Ruiz--Lapuente et al. 1995; Ruiz--Lapuente et al. 1997),
 as we have mentioned above, 
is linked to the time the WD has cooled before reaching the point of 
explosion. This affects the amount of neutronized material synthesized at the 
center of the exploded WD and the overall nucleosynthesis. Another possible 
effect is related to the composition of the accreted material added to the 
initial WD till the explosion occurs. If the WD gains mass by accreting H 
from a non-WD companion and burning it to He and C+O, and the initial mass of 
the WD is smaller (although not by a large factor) in an old population, the 
star needs to add up more material before reaching the explosion in old than 
in young systems. This creates differences in the composition of the outer 
layers of systems belonging to different populations and that could 
be reflected in the final brightness--rate of decline relation (by an opacity 
effect). We favor, however the first (cooling) effect, since the typical age 
of the binary system when it reaches explosion is quite long, in the most 
likely SNIa picture. For the WDs exploding as SNe Ia, typical ages $\sim 
5\times 10^{9}\ yr$ and a small mass variation between $z = 0$ and $z = 0.8$ 
are found. 

In the 3--D hydrodynamic simulations of WD explosions
 by Hillebrandt et al. (2000),
the family of SNe Ia seems to arise from different conditions at the
start of the thermonuclear burning of the C+O WD,
 in particular, the number/location
of spots where
the ignition starts and the density at which the ignition takes place. 
Different results for the nucleosynthesis are obtained: 
a more complicated topology of the initial flame seems to lead to 
higher Ni--production and, consequently, more powerful and brighter explosions
(Reinecke, Hillebrandt \& Niemeyer 2002). Light curves 
for this consistent set of  models reproduce well the observations (see 
Blinnikov at this conference; Blinnikov \& Sorokina 2000). As the 
initial conditions in the convective
core of the WD in the pre--explosion stage are very critical
for the development of the burning (Garcia--Senz \& Woosley 1995), 
it seems natural to think 
that those factors determining the outcome 
are linked to the cooling undergone by the WD prior to explosion.

\section{Modeling of Type Ia SN  at various z}

\begin{figure}
\input epsf2
\bigskip
\bigskip
\bigskip
\epsfxsize=180pt
\epsfbox{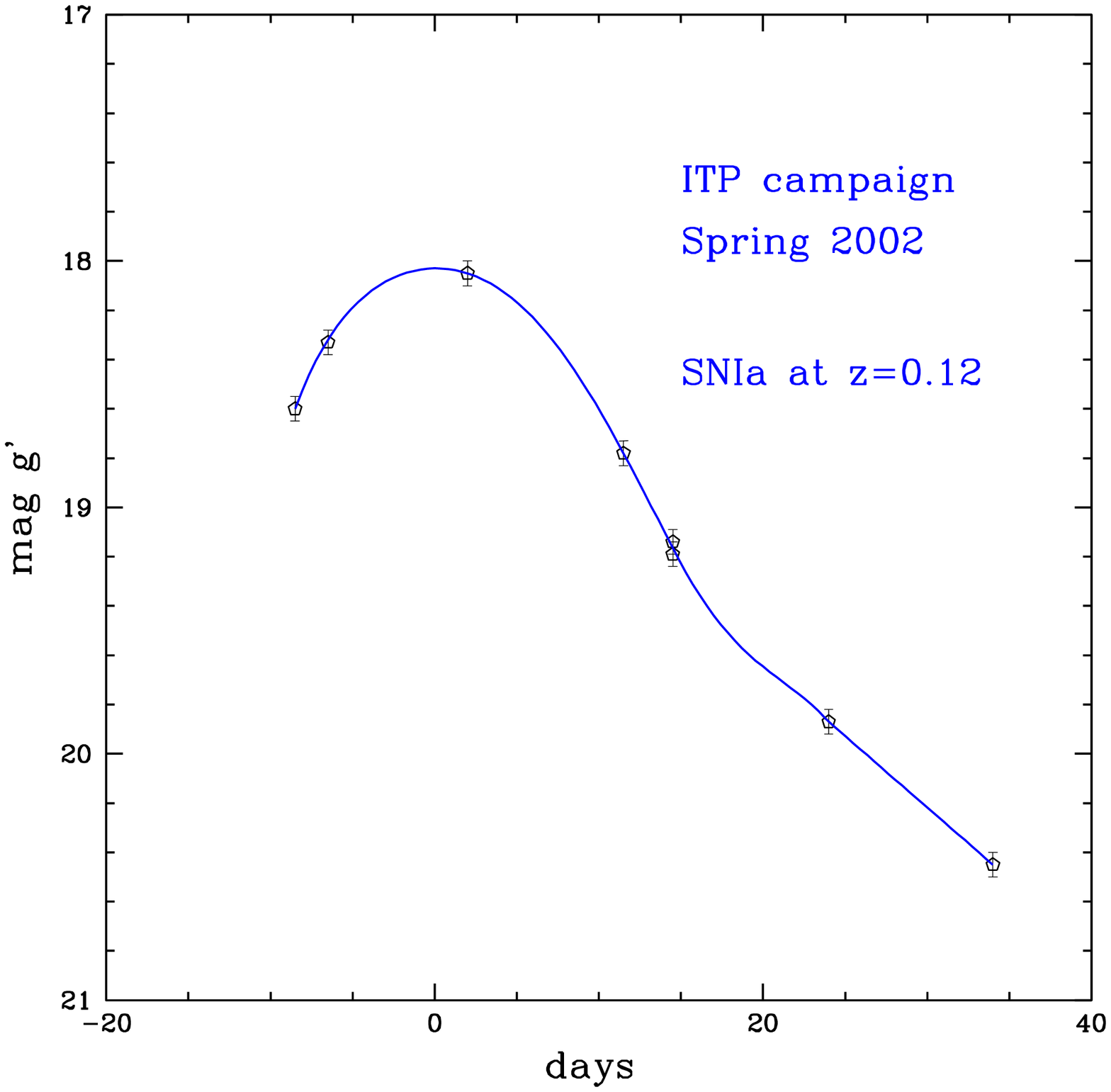}
\epsfxsize=220pt
\epsfbox{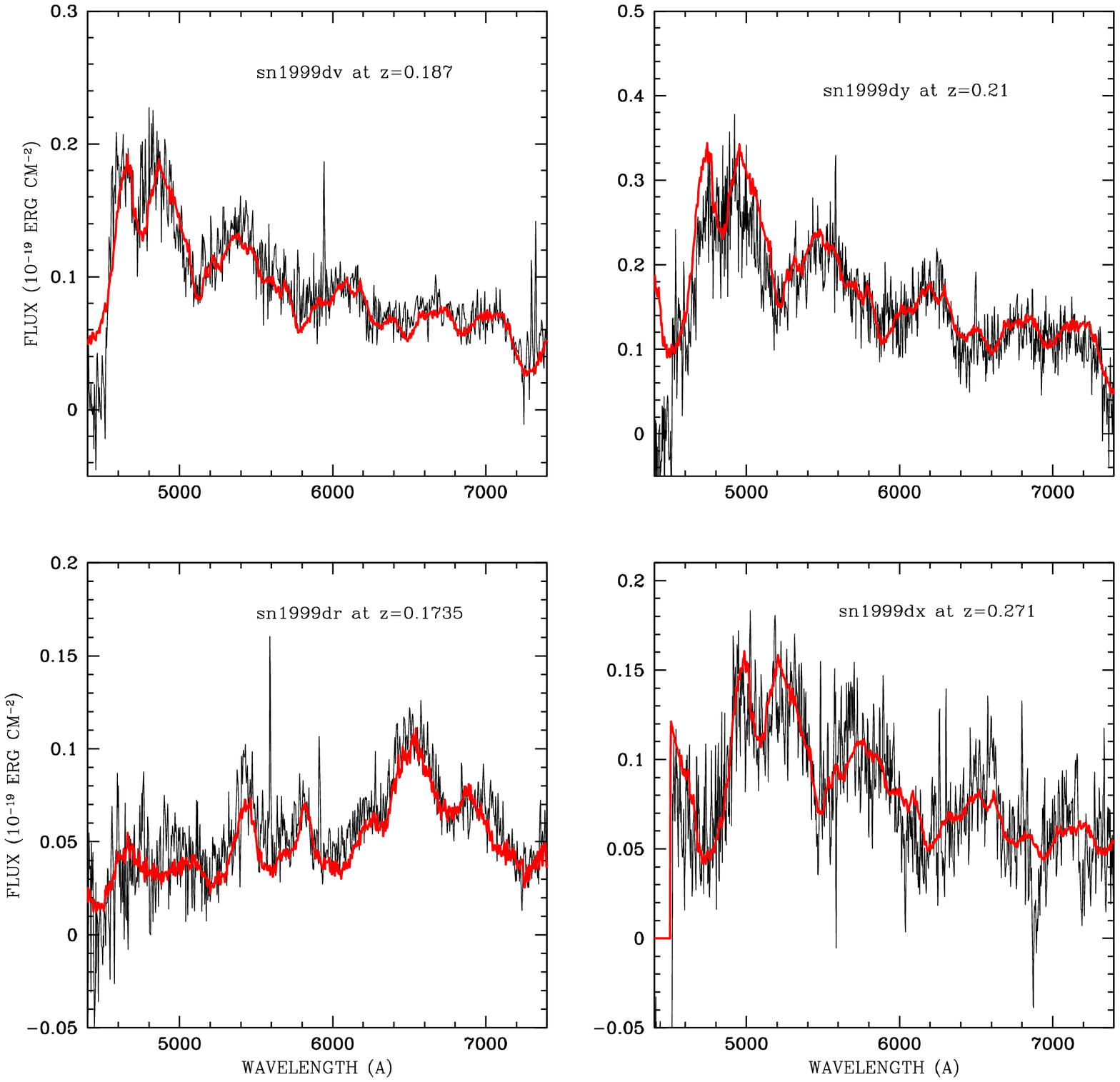}
\vspace{6mm}
\caption{
SNe Ia at intermediate $z$  found 
in campaigns by the ESCC.
The lower panel shows the
spectra of four SNe Ia discovered during the 1999 campaigns 
(Hardin et al. 1999; Pain et al. 2002).
The upper panel shows
the good sampling of the light curves obtained in the spring 2002 
campaign which was conducted within the International 
Time Programme {\it  $\Omega$ and $\Lambda$
through SNe Ia and the Physics of Supernova Explosions}, carried out 
at the telescopes of the ENO at La Palma (Ruiz--Lapuente et al. 2002a). 
}

\end{figure}

The discovery of the accelerating expansion of the Universe was mostly based 
on observations of SNe Ia at $z\sim 0.5$. The currently preferred values for
the matter density $\Omega_{M} \sim 1/3$ and the dark energy density
$\Omega_{\Lambda} \sim 2/3$ imply that the expansion began to accelerate
at $0.5 < z < 1.0$, that is between 4.3 Gyr and 6.7 Gyr ago.      
There was a gap centered around $z\sim 0.2$ between the local sample and 
the high--redshift one, which we aimed to fill.
To help in those cosmological studies  
we\footnote{Members 
of the European Supernova Cosmology Consortium  involved 
in the first campaigns  are P. Astier, C. Balland, G. Blanc, S. Blinnikov, 
R. S. Ellis, S. Fabbro, G. Garavini,  A, Goobar, D. Hardin, I. Hook, 
M. Irwin, R. G. McMahon, J. M\'endez, M. Mouchet, A. Mourao,
 S. Nobili, J. Rich, P. 
Ruiz--Lapuente, K Schahmaneche, E. Sorokina, R. Taillet, N. Walton.}
have carried out observations and done the modeling of SNe Ia in a restricted 
$z$ interval chosen to be $z\sim 0.15-0.35$. We are placing 
our emphasis on investigating systematic effects and 
 comparing theoretical and observed light curves by modeling 
the explosions of WDs (hydrodynamics, nucleosynthesis) and the corresponding 
light curves.

\begin{table}[htp]
\begin{center}
\caption{SNeIa targeted in the International Time Programme}

\begin{tabular}{llll}
\hline
         SN Targets &  z   &  SN  & Subprogramme \\
\hline
  Nearby & $< 0.016$  & SN 2002bo, SN 2002dj, SN 2002er & 
 Intensive follow up\\
  Medium-z & 0.1-0.3 & Completion of a sample of 20 SNe Ia & Cosmology \\
\hline
\end{tabular}
\end{center}
\end{table}

\begin{figure}
\input epsf2
\epsfxsize=210pt
\epsfbox{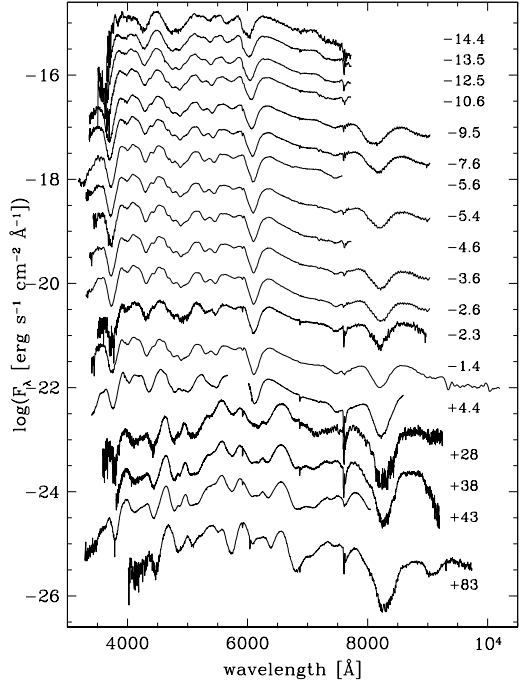}
\vspace{6mm}
\caption{SN 2002bo in NGC in NGC 3190 (Benetti et al. 2002).
  Intensive spectroscopic follow--up
  within the programme {\it The Physics of Type Ia Supernova Explosions}
 and the ITP on  {\it $\Omega$ and $\Lambda$  from SNe Ia and the Physics 
 of Supernova Explosions}.}
\end{figure}

\begin{figure}
\input epsf2
\epsfxsize=210pt
\epsfbox{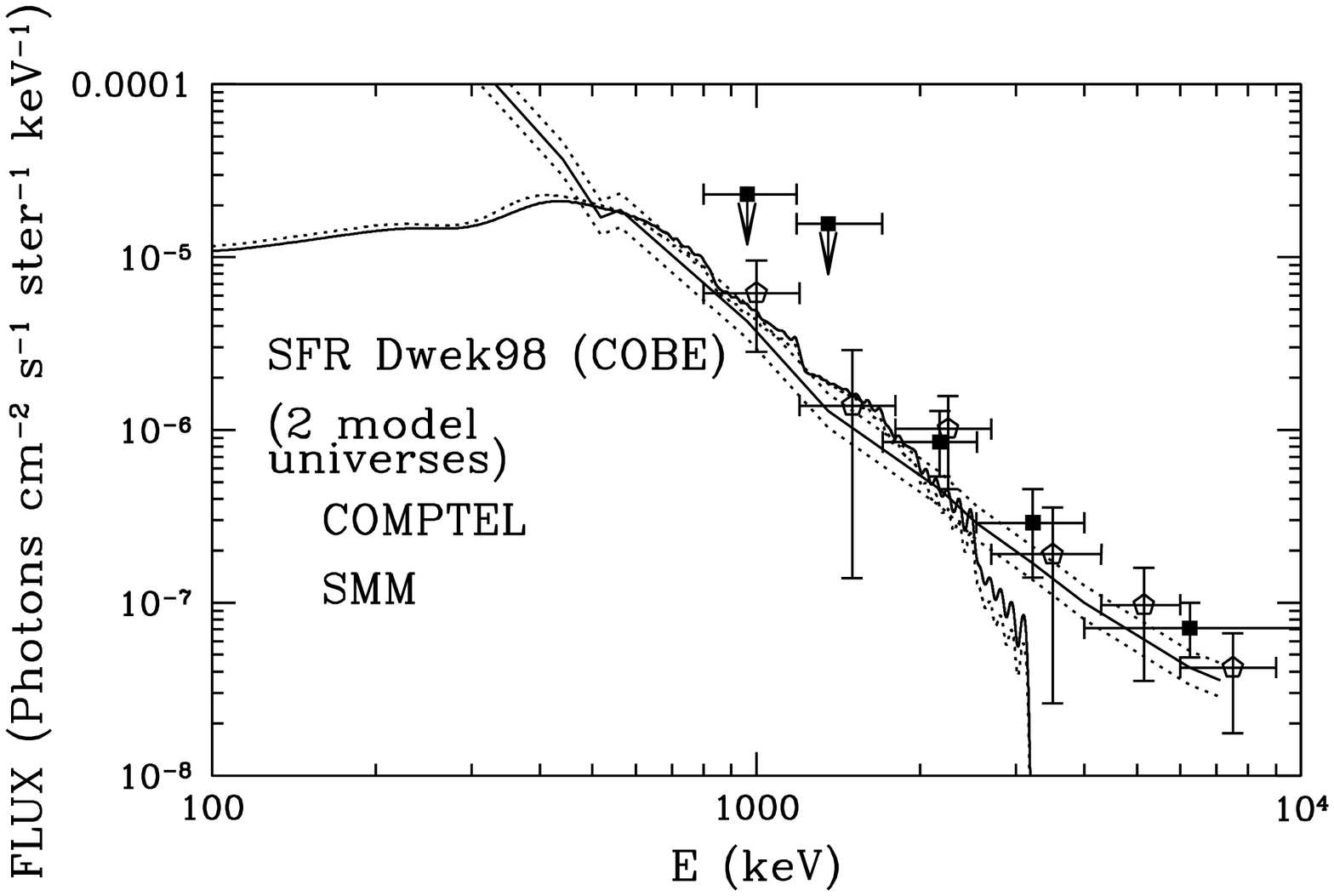}
\vspace{6mm}
\caption{
Type Ia SNe made out of stars at various $z$ can be identified as responsible 
for the extragalactic $\gamma$--ray background in the MeV range 
(Ruiz--Lapuente, Cass\'e \& Vangioni--Flam 2000). 
The background light produced by SN basically arises from SN up to $z\sim 1$. 
SNe Ia rates compatible with the results from optical supernova  searches 
give a background emission in the MeV range that can indeed explain the 
extragalactic emission measured by COMPTEL and SMM. }
\end{figure}

 As it had been shown by 
previous campaigns, within that redshift interval observations with 4.2m--2.5m 
telescopes can still be complete enough to explore possible deviations of the 
peak luminosity--light curve shape relationship from the local one, including 
the faintest SNe Ia, check for correlations of SNe Ia properties with galaxy 
type and with location inside a given galaxy, and measure intergalactic 
extinction. In addition, improved accuracy on the evolution of the SNe Ia rate 
along $z$ should result. Finally, the interval being well within the 
acceleration era, the SNe Ia observations can contribute valuable information 
on $w_{X}(z)$ itself.

The searches of SNe Ia at redshifts centered around $z\sim 0.2$
 used the {Wide Field Camera} (WFC) on the 2.5m INT. That 
telescope was used as well for the follow up. Spectroscopy was obtained at
the 4.2m WHT allowing a classification of the supernovae. 
 The first campaign took place in March--April 1999 and led to the
detection of 19 SNe Ia candidates from which 2 were identified as SNe Ia 
(Hardin et al. 2000). 
This first INT run
searched 12 sq degrees in R and B with magnitude limit 23 in B and 22 in 
R. Rates of SNeIa at $z = 0.3$, as determined by Pain et al. (1996) 
are 20  SNe Ia degree$^{-1}$ night$^{-1}$ 0.5 mag$^{-1}$.
 For 1 night and integrating for 6 half
magnitudes this implies rates of 0.33 SNe Ia sq deg$^{-1}$night $^{-1}$ 
between mag 18
and 21. The discoveries made seemed consistent with early 
findings (Pain et al. 1996).
 The September--October INT run obtained 15 candidates 
by searching 30 sq degree in $g'$ up to a magnitude limit $g'\sim 22.5$. The 
mean $z$ of the 8 confirmed SNe Ia is $\langle z\rangle = 0.3$
 (Hardin et al. 2000). 
During  March--April 2000 we developed our campaign in combination with 
EROS, aimed to the discovery of SNe Ia at low $z$. 10 SNe Ia were 
identified at 
a $\langle z\rangle =0.05$. The most recent campaigns  
(Ruiz--Lapuente et al. 2002a) have gathered a  sample of 10
SNe Ia at $\langle z\rangle =0.2$.

 In parallel with the empirical exploration of SNeIa properties along z, we
 have undertaken the detailed photometric and spectroscopic study of a 
 sample of nearby SNe Ia discovered before they reached maximum light. 
 Within this frame, 
 intensive studies of three SNeIa in the local 
Universe were  done\footnote{Work done in collaboration
 with the Asiago and  ESO observatories within the European Research
 and Training Network on 
 Physics of SNe Ia. http://www.mpa-garching.mpg.de/rtn.}. 
 Figure 8 shows the intensive spectroscopic follow up done for
 SN 2002bo (Benetti et al. 2002). 

Questions related to the physical spread of the SNe Ia family, will ultimately 
point towards the way the white dwarf accretes mass and reaches the 
state immediately preceding the explosion. The nature of the binary system in 
which SNe Ia take place is still unknown. A program to 
 look for the companion star in the two historical SNe Ia 
in the Galaxy have yielded some constraining
 results (Ruiz--Lapuente et al. 2002b).

\begin{figure}
\input epsf2
\bigskip
\bigskip
\bigskip
\epsfxsize=300pt
\epsfbox{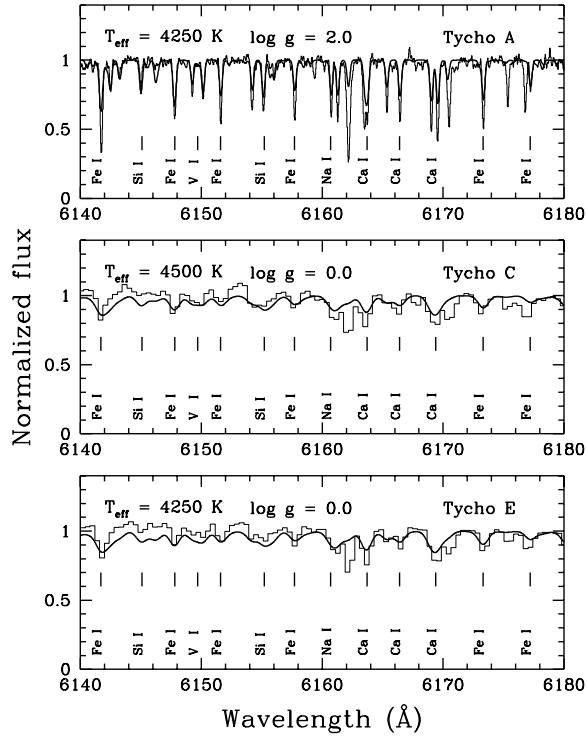}
\vspace{6mm}{ }
\caption[]{ 
Calculated synthetic spectra compared with the observed spectra of
the SN companion candidates near the center of SN 1572.
These are the closest red giants and 
supergiants to the center of the explosion. Their surface gravity goes
from log g=2 for Tycho A to log g=0 for C and E. The effective 
temperatures for those stars are similar. They
are in the range 4200--4500 K. Model atmospheres with solar chemical 
abundances give a good account of the spectra. 
The overall spectral comparison allows us to
exclude overabundances of 
the Fe--peak elements. Moreover, the stars show no enhancement of 
iron--peak elements versus intermediate--mass elements in the
spectra. Synthetic spectra are shown with bold continuous lines. 
Those comparisons rule out some systems as  progenitors 
of the Galactic SNeIa (Ruiz--Lapuente et al. 2002b).}
\label{fig1}
\end{figure}

\section{Stars giving Type Ia SNe }

\subsection{SNeIa efficiencies and progenitors}

The high--$z$ supernova searches not only allow to trace the expansion
history of the Universe and provide a new picture of its matter--energy
contents, but they also throw new light on a number of important 
astrophysical questions. 

The SNe Ia production efficiency will first give us clues as to 
the nature of the so far elusive SNe Ia binary progenitor systems. 

We know that the exploding star giving rise to a Type Ia SN is a 
carbon--oxygen white dwarf (C+O WD) and that there is no compact object left 
in the explosion. Stellar evolution arguments tell us as well that those 
explosions take place  in binary systems. Up to recently, there was no clear 
evidence favoring any particular kind of binary system as responsible for the 
Type Ia phenomenon. Two candidate systems were proposed. One possibility is 
that SNe Ia arise from a binary made of a pair of C+O WDs  
that merge as their orbit shrinks due to the emission of gravitational wave 
radiation. This system is referred  to {\it double degenerate}.
The other candidate system is a WD which accretes material from a Roche--lobe 
overfilling non-WD companion (WD plus Roche--lobe filling subgiant, giant or
main sequence star). 
The system contains only a WD and therefore
is a {\it single--degenerate system}
 (see Branch et al. 1995; Ruiz--Lapuente et al. 1997; Livio 2000 
and references therein).

If we evaluate the number of SNeIa exploding per unit comoving volume in 
redshift space $\Re_{Ia}(z)$ relative to the mass going into forming stars in 
the Universe $\dot \rho_{*}(z)$, we have a quantity that gives the 
efficiency of stars in producing SNe Ia along $z$ space. We can express this 
efficiency as ${\cal E}_{SNeIa}(z)$:

\begin{equation}
{\cal E}_{SNeIa}(z) = \Re_{Ia}(z)\ yr^{-1}\ M\!pc^{-3}\ /\ 
\dot \rho_{*}(z)\ M_{\odot}\ yr^{-1}\ M\!pc^{-3}
\end{equation}

\noindent
The quantity ${\cal E}_{SNeIa}(z)$ above is just the number of SNe Ia per unit 
mass spent in forming stars at a given $z$, and it is independent from the 
cosmological model assumed. The absolute values of ${\cal E}_{SNeIa}(z)$ 
reflect the abundance of progenitor systems (and thus the range of initial 
conditions leading to SNe Ia). It also reflects the evolutionary time scale 
(from birth to explosion) of the progenitor systems, together with other 
possible evolutionary effects. The 
single degenerate systems (made of a WD accreting from a non--WD companion)
have evolutionary time scales of the order of a few Gyr whereas the double 
degenerate systems (coming from the merging of two WDs) have time scales of 
the order of a few hundred million years only.

\begin{table}
\caption{Observations of SN 1572}
\begin{center}
\renewcommand{\arraystretch}{1.4}
\setlength\tabcolsep{5pt}
\begin{tabular}{lllllll}
\hline\noalign{\smallskip}
Run  & Rd ($'$)$^{1}$  & m$_{R}$ $^{2}$  & Telescope $^{3}$  & R
  &  Spec Range (A) & stellar types  \\
\noalign{\smallskip}
\hline
\noalign{\smallskip}
 (1) & 0.7 & 14 & WHT (UES) & 50,000 & 4000--7100 & red giant \\
 (2) & 0.7 & 23 & WHT (ISIS) & 15,000  & 4600-7500  & red giants to WD \\
 (3) & 0.7 & 23 & Keck (ESI) & 7000 & 4000--10000  & as above \\
\hline
\end{tabular}
\end{center}
\label{Tab1a}
\end{table}
\begin{table}
\caption{Observations of SN 1006}
\begin{center}
\renewcommand{\arraystretch}{1.4}
\setlength\tabcolsep{5pt}
\begin{tabular}{lllllll}
\hline\noalign{\smallskip}
Run  & Rd ($'$)$^{1}$  &  m$_{R}$ $^{2}$& Telescope $^{3}$ & R
  & Spec range (A) & stellar types \\
\noalign{\smallskip}
\hline
\noalign{\smallskip}
 (1) & 5 & 13 & NTT (EMMI) & 10,000 & 3950--7660 & red giants \\
 (2) & 5 & 15 & VLT (UVES) & 50,000 & 3500--9000  &  all types \\

\hline
\end{tabular}
\end{center}
\label{Tab1a}
\noindent$^{1}$ Radius of the search

\noindent$^{2}$ Limiting magnitude 

\noindent$^{3}$ Telescopes(Instrumentation) 

\end{table}

Despite the uncertainties inevitably involved, the high--$z$ searches have 
nevertheless brought some crucial information. The measurements of the cosmic 
evolution of the SNe Ia rate now extend up to $z\sim 
0.55$ (Pain et al. 1996; 2002).
The mean value of ${\cal E}_{SNeIa}(z)$ over the redshift 
interval $0 < z < 0.55$ corresponds to 1 SNeIa per 
$709^{+200}_{-157}\ M_{\odot}$ spent in forming stars. The low--redshift 
efficiencies ($z\simeq 0$) correspond to 1 SNeIa per $\sim 900\ M_{\odot}$ 
going into star formation, reaching 1 SNIa per $\sim 700\ M_{\odot}$ at 
$z\simeq 0.55$. The efficiency of binaries in ending as SNeIa 
is 6\% of the stars 
between $3-9\ M_{\odot}$. Therefeore, the channels that give
rise to such high efficiency in giving SNeIa have to be wide.

A more decisive test comes from the fact that in SNeIa
 from single degenerate systems,
 we can expect detecting motion of the non--WD companion 
imparted when the systems disrupts, and that could be the key to the 
final identification of the SNe Ia progenitor 
systems (Ruiz--Lapuente 1997; Marietta, Burrows \& Fryxell 2000; 
Canal, Mendez \& Ruiz--Lapuente 2001). Since 1997 (Ruiz--Lapuente 1997)
 searches for the moving 
companions of SN 1572 and SN 1006 have been performed. High resolution
spectra of the stars within the radius of the historical remnants have
been taken (see Table II and III) and their
 radial velocities have been measured.  
The spectra of these stars 
 have been modeled to test for contamination by
the supernova explosion. 
 The lack of significant radial velocities and 
contamination from the SN explosion in these stars 
 (see results for SN 1572 in Ruiz--Lapuente et al. 2002b) puts 
severe constraints on the progenitors. Ultimately, it comes out that
going back to the Galaxy is needed to tie up the loose ends.

\section{Prospects for dark energy measurements along z} 
 
If the SNe Ia results are correct, as they seem to be, our Universe is 
accelerated by a component which has a negative pressure. It has been 
discussed that it is  
possible to test gravitation theories by going to a higher level of analysis 
using Type Ia SNe (Weller \& Albrecht 2001).  Possibilities 
include testing SUGRA potentials, as 
discussed in many recent papers. The solution corresponding to the 
cosmological constant could be discriminated against quintessence or other 
scalar field models (Huterer \& Turner 1999; Steinhard et 
al. 1998). Several experiments are under way
 to shed light on the nature of the 
dominant component of our Universe by observing a large number of SNe Ia 
up to $z = 1.7$. Strategies for a better discrimination of the dark 
energy are discussed by Huterer \& Turner (2000) .

  The observational  determination of 
 {\it w} and its variation with z has already started: while
the Supernova Cosmology Project is enlarging the sample of high--z SNe Ia, 
the SNfactory\footnote{http://SNFactory.lbl.gov} 
will provide an anchoring 
in the low--z domain with hundreds of SNe Ia below z $=$0.2. 
ESSENCE\footnote{http://www.ctio.noao.edu/essence} plans as well to follow
 $\sim$ 200 SNe Ia in the z interval 0.15--0.75 over a five year period.
 About a 
thousand   SNe Ia discoveries  are expected to come from the
 CFHTLS\footnote{http://www.cfht.hawaii.edu/Science/CFHTLS}
during the next 5 years, with preliminary results 
on the equation of state, while in the very high-z range the GOODS 
{\it HST} Treasury
 Program\footnote{http://www.stsci.edu/ftp/science/goods} can 
bring supernovae to test the epoch of the deceleration of the Universe
(z $>$ 1). To add to this the intermediate z searches 
and the very nearby ones\footnote{http://www.mpa-garching.mpg.de/~rtn} should
greatly enhance the knowledge of SNe Ia and reinforce their cosmological use.

Ultimately, to unveil the nature of dark energy, 
ground--based programmes are limited 
in accuracy and scope. It seems unavoidable to go to a fully devoted mission
from space, such as SNAP\footnote{http://snap.lbl.gov} 
to achieve an improved level of accuracy 
in the cosmological measurements, and to be able to discriminate among 
possible candidates to dark energy.

From all the above one sees that the various 
 steps in the road to test the cosmological implications of SNe Ia 
will be taken. Thus, we expect the next years to
bring definitely an understanding of what lies   
behind the observed acceleration of the expansion of the Universe.

Work on Supernovae and Cosmology by the author is supported in part
by grant AYA2000--0983 and RTN2-2001--0037. I thank my collaborators
for their inspiring contributions to the work reviewed here.

\end{article}

\begin{thebibliography}{}

\bibitem{aldering99}Aldering, G., Knop,R., \& Nugent, P., {\it AJ} {119}, 210
 (2000).
\bibitem{aguirre99}Aguirre, {\it ApJ} {\bf 525}, 583 (1999).
\bibitem{barbon99}Barbon, R., Buondi, V. Cappellaro, E. \& Turatto, M.
 {\it A \&A S} {\bf 139}, 531 (1999).
\bibitem{benetti02}Benetti, S. et al., in preparation (2002).
\bibitem{blinnikov00}Blinnikov, S.I., \& Sorokina, E.I., {\it A\&A} {\bf 356},
L30 (2000).
\bibitem{branch95}Branch, D., Livio, M., Yungelson, L.R., Boffi, F.R., \& 
Baron, E., {\it PASP} {\bf 107}, 1019 (1995).
\bibitem{branch96}Branch, D., Romanishin, W., \& Baron, E., {\it ApJ} 
{\bf 465}, 73 (1996).
\bibitem{canal00}Canal, R., M\'endez, J., \& Ruiz--Lapuente, P., {\it ApJ}
 {\bf 550}, L53 (2001).
\bibitem{dwek98}Dwek, E., et al., {\it ApJ} {\bf 508}, 106 (1998). 
\bibitem{sullivan00}Ellis, R., \& Sullivan, M., {\it New Cosmological Data 
and the Values of the Fundamental Parameters, IAU Symp 201} {}, in press 
(2000).
\bibitem{farrah02}Farrah, D., Meikle, W.P.S., Clements, D., 
Rowan--Robinson, M., \& 
Mattila, S., {\it MNRAS} {\bf 336}, L17 (2002).
\bibitem{filip92a}Filippenko, A.V., et al., {\it ApJ} {\bf 384}, L15 
(1992).
\bibitem{filip92b}Filippenko, A.V., et al., {\it  AJ} {\bf 104}, 1543 
(1992).
\bibitem{filippenko98}Filippenko, A.V., \& Riess, A.G., {\it Phys. Rep.} 
{\bf 307}, 31 (1998).
\bibitem{garcia95} Garcia--Senz, D. \& Woosley, S.E.   {\it ApJ} {\bf 454}, 895
(1995).
\bibitem{goldhaber01}Goldhaber, G., et al., {\it ApJ} {\bf 558}, 359  (2001).
\bibitem{goobar95}Goobar, A.,\& Perlmutter, S., {\it ApJ} {\bf 450}, 14 (1995).
\bibitem{hamuy95}Hamuy, M., et al., {\it AJ} {\bf 109}, 1 (1995).   
\bibitem{hamuy96a}Hamuy, M., et al., {\it  AJ} {\bf 112}, 2399 (1996).
\bibitem{hamuy96b}Hamuy, M., et al., {\it AJ} {\bf 112}, 2391 (1996).
\bibitem{hamuy99}Hamuy, M., \& Pinto, P.A., {\it AJ} {\bf 117}, 1185 (1999).
\bibitem{hardin001}Hardin, D., et al., talk given at Moriond 
 conference on {\it Energy Densities in the Universe}.  
in press (2000).
\bibitem{reinecke00}Hillebrandt, W., Niemeyer, J.C., \& Reinecke, M., 
{\it Cosmic Explosions, 10th Ann. Astrophys. Conf. Univ. Maryland}, 2000,  
p53.
\bibitem{huterer99}Huterer, T., \& Turner, M.S., {\it Phys. Rev. D} {\bf 60},  
1301 (1999).
\bibitem{huterer00}Huterer, T., \& Turner, M.S., {\it Phys. Rev. D} {\bf 64},  
3527 (2001).
\bibitem{ivanov00}Ivanov, V.D., Hamuy, M., \& Pinto, P.A., {\it ApJ} 
{\bf 542}, 588 (2000).
\bibitem{leibund93}Leibundgut, B., et al., {\it  AJ} {\bf 105}, 301 (1993).
\bibitem{leibundgut00}Leibundgut, B., {\it New Cosmological Data and the 
Values of the Fundamental Parameters, IAU Symp 201} {}, in press (2000).
\bibitem{livio00}Livio, M.{\it Type Ia Supernovae, Theory and Cosmology}, 
ed. J.C. Niemeyer \& J.W. Truran, Cambridge University Press, 33 (2000) 
\bibitem{marietta00}Marietta, E., Burrows, A., \& Fryxell, B., {\it ApJS}, 
{\bf 128}, 615 (2000).
\bibitem{maza94}Maza, J., et al., {\it ApJ} {\bf 424}, L107 (1994).
\bibitem{nugent98}Nugent, P., Phillips, M.M., Baron, E., Branch, D., \&
Hauschildt, P., {\it  ApJ} {\bf 455}, L147 (1998).
\bibitem{pain96}Pain, R., et al., {\it ApJ} {\bf 473}, 356 (1996).
\bibitem{pain02}Pain, R.,  et al., {\it ApJ} {\bf 577}, 120 (2002).
\bibitem{pain02}Pain, R. et al. 2002, in preparation. 
\bibitem{perlmutter99}Perlmutter, S., 
et al. (The Supernova Cosmology Project), 
{\it ApJ} {\bf 517}, 565 (1999).
\bibitem{phillips93}Phillips M.M., {\it ApJ} {\bf 413}, L105 (1993).
\bibitem{phillips99}Phillips, M.M., et al., {\it AJ} {\bf 118}, 1766 (1999).   
\bibitem{pskovskii77}Pskovskii, Y.P., {\it Soviet Astron.} {\bf 21}, 675 
(1977).
\bibitem{quimby02}Quimby, R. et al. (The Supernova Cosmology Project),
{\it BAAS} {\bf 201}, 2305 (2002).
\bibitem{hillebrandt02}Reinecke, M., Hillebrandt, W. \& Niemeyer, J.C.,
{\it A \& A } {\bf 391}, 1167 (2002).
\bibitem{riess95a}Riess, A.G., Press, W.H., \& Kirshner, R.P., {\it ApJ} 
{\bf  438}, L17 (1995a).
\bibitem{riess95b}Riess, A.G., Press, W.H., \& Kirshner, R.P., {\it  ApJ} 
{\bf 438}, L17 (1995b).
\bibitem{riess98}Riess, A., et al., {\it AJ} {\bf 116}, 1009 (1998).
\bibitem{riess00}Riess, A., et al., {\it AJ} {\bf 118}, 2668 (1999).
\bibitem{riess2000}Riess, A., et al., {\it ApJ} {\bf 536}, 62 (2000).
\bibitem{riess01}Riess, A.G. et al. ApJ 560,{\bf 49} (2001). 
\bibitem{rowanrobinson02} Rowan--Robinson, M. 
{\it MNRAS} {\bf 332}, 352 (2002).
\bibitem{ruiz--lapuente95b}Ruiz--Lapuente, P., Burkert, A., \& Canal, R.,
{\it ApJ} {\bf 447}, L69 (1995).
\bibitem{ruiz--lapuente95}Ruiz--Lapuente, P., {\it From Quantum Fluctuations 
to Cosmological Structures, ASP Conf. Ser.} {\bf 126}, 207 (1997).
\bibitem{ruiz--lapuente97}Ruiz--Lapuente, P., {\it Science} {\bf 276}, 1813 
(1997).
\bibitem{ruiz--lapuente97a}Ruiz--Lapuente, P., Canal, R., \& Burkert, A., {\it 
Thermonuclear Supernovae}, Dordrecht: Kluwer, 1997, pp. 205--230.
\bibitem{ruiz--lapuente00}Ruiz--Lapuente, P., Cass\'e, M., \& Vangioni--Flam, 
E., {\it ApJ} {\bf 549}, 483 (2000).
\bibitem{ruiz--lapuente0002}Ruiz--Lapuente, P. et al. 
  ITP2002 on {\it $\Omega$ and $\Lambda$  from SNe Ia and the Physics 
 of Supernova Explosions} (2002a).
\bibitem{ruiz--lapuente0002}Ruiz--Lapuente, P., Comeron, F., 
Smartt, S., Kurucz, R., Mendez, J., Canal, R., Filippenko, A. \&
Chornock, R. 2002, {\it From Twilight to Highlight: The Physics 
of Supernovae}, ed.
W. Hillebrandt \& B. Leibundgut (Springer Verlag, Berlin) 2002b. 
\bibitem{sullivan02} Sullivan et al. (The Supernova Cosmology Project),
 {\it MNRAS}, in press (2002). 
\bibitem{steinhardt00}Steinhardt, P.J., et al., {\it Phys. Rev. D} {\bf 59},
 (1999).
\bibitem{weller00}Weller, \& Albrecht,  {Phys. Rev. D} {\bf 86} 1939 (2001).


\end{thebibliography}
\end{document}